\documentclass{article}
\usepackage[utf8]{inputenc}
\usepackage[english]{babel}
\usepackage{amsmath}
\usepackage{amssymb}
\usepackage{amsthm}
\usepackage{xurl}
\usepackage{hyperref}
\usepackage{tabularx}

\theoremstyle{theorem}
\newtheorem{theorem}{Theorem}[section]

\theoremstyle{definition}
\newtheorem{definition}[theorem]{Definition}

\theoremstyle{remark}

\title{Hashing geographical point data using the space-filling H-curve}
\author{
    Igor V. Netay
    \thanks{Joint Stock "Research and production company ``Kryptonite"}
    \thanks{Institute for Information Transmission Problems, Russian Academy of Sciences}
    \href{mailto:i.netay@kryptonite.ru}{i.netay@kryptonite.ru}
}

\date{}

\begin{document}
\maketitle

\begin{abstract}
    We construct geohashing procedure based on using of space-filling H-curve.
    This curve provides a way to construct geohash with less
    computations than the construction based on usage of Hilbert curve.
    At the same time, H-curve has better clustering properties.
\end{abstract}


\section*{Introduction}

A space-filling curve (below~--- SFC for shortness) of dimension~$d$
is a map from~$[0,1]$ into~$\mathbb{R}^d$
such that the image contains an open non-empty set or, equivalently, some
cube~$[a_1,b_1]\times\ldots\times[a_d,b_d]$ for~$a_i<b_i$, $i=1,\ldots,d$.
There are lots of space-filling curves, many of them can be found in~\cite{16Curves}.

\textit{Fractal space-filling curves} are self-similar curves, i.\,e. they exhibit
similar patterns at increasingly small scales.
This similarity is called \textit{unfolding symmetry}.
If the symmetry is a composition of unfolding and an isometry ones,
the curve is called \textit{affine self-similar}.

Affine self-similar curves play an important role in practice, because they can be easily
constructed and provide a way to construct locality-preserving mapping from multidimensional
data into one-dimensional space.
Examples of applications include geo-information (see~\cite{GIS}) systems and 
database indexes (see~\cite{908985},~\cite{SecKey}).

The simplest known curve is called Morton curve (also known as ``Z-curve'', see~\cite{Z}).
Its improvement by usage of Grey coding was proposed in~\cite{Gray}.
The more compoex and most widely used curve is called Hilbert curve (see~\cite{Hilbert}).

In \cite{vukovic2016hilbert} it was shown that Hilbert curve has superior hashing
properties to Z-curve.
At the same time, it is more computationally  complex.
Here we introduce geohashing procedure based on usage of H-curve introduced in~\cite{HCurve}.

We compare H-curve with Z-curve and Hilbert curve in sense of locality preservance and
in the context of hashing algorithm performance.

It will be shown that the implementation based on~\cite{HCurveGit}
gives significantly better performance than Hilbert curve.
The performance of the curve point evaluation is $4$--$8$ times faster for
H-curve than for Hilbert curve (see~\cite[Table~1]{HCurve}).
In the case of geohash construction the performance is also significantly
better, see~\S\ref{sec:bench}.

We use implementation of~\cite{vukovic2016hilbert} on \texttt{golang} as a base.
The implementation of H-curve is made on \texttt{C++}, so they are linked together,
and all the benchmarks and distance comparisons are made with the same code
modified only to add H-curve and link it against the H-curve implementation
designed previously on \texttt{C} and \texttt{Haskell}.
That code was translated to~\texttt{C++} in order to combine sources via~\texttt{C API}.

The source code of modified benchmarks and metrics and translated H-curve with 
build scripts is available on GitFlic (\cite{HGHGit}).
It is based on source code of~\cite{vukovic2016hilbert} available on~\cite{VukGit}.

\section{Curves comparison}
\label{sec:m}

Effectiveness of space-filling curves usage is compared in terms of so-called
clustering property being a measure of ``preserving locality''.
It is well known that fractal space-filling curves are surjective continuous
mapping from segment~$[0, 1]$ to the $d$-dimensional cubes~$[0, 1]^d$.
These mapping are never (except~$d=1$) injective and homeomorphic.

For practical issues, finite iteration of curve constructions are applied:
$n$-th iterations corresponds to subdivision of a unit $d$-dimensional cube
into $2^{nd}$ subcubes ($2^n$ along each axis).
This cubes form a lattice inside initial unit cube.
A finite step for fixed dimension~$d$ is a linear ordering on $2^{nd}$ unit cubes.
Here $n$ is called \textit{subdivision granularity}.

Suppose we construct index in a database from $d$ features being integral
of floating point numbers.
This index is usually constructed by some hashing procedure like computation of SFC-index.
In this case the value of hash is the number of point (or small cube) for
subcubes traversal induced by the curve construction.

One of approaches to measure ``quality'' of SFC is to calculate average number
of consequent blocks in indices for geometrical queries, i.\,e.~subsets of points
in the unit cube in data space lying inside some geometrical shape in the real space spanned
by lattice points representing data space.

We are interested in database rectangular queries being simply subsets of
the lattice points inside some many-dimensional cube.

Let us recall some definitions.

\begin{definition}
    A subset~$p\subseteq q$ of a query is called a \textit{cluster} with respect to a SFC~$\omega$
    if it is a maximal subset such that the points (or cells) of~$p$ are numbered sequentially by~$\omega$.
    We denote the number of clusters in~$q$ by~$c_q(\omega)$.
\end{definition}

\begin{definition}
    A \textit{clustering property} of a SFC~$\omega$ with respect to a
    (maybe parametric) class of queries~$\mathcal{Q}$
    is the average number~$c_{\mathcal{Q}}(\omega)$ of clusters
    in~$q\in \mathcal{Q}$ (or the limits/asymptotics of cluster number
    as a function in the parameters if exist).
\end{definition}

It turns out that Hilbert curve and H-curve have very close clustering property (see~\cite[Table~2]{HCurve}).

Here we compare space-filling curves in terms of Levenstein distance between 
hashes (also known as ``edit distance'').
It is more suitable metric for applications needing the clustering of database queries 
and for some other applications like URL shortening.
For instance, Hilbert curve gives less Levenstein distance than Z-curve in~$\approx 53\%$
of cases (see~\cite{vukovic2016hilbert}).
It turns out that H-curve gives shorter distance in~$\approx 74\%$ cases than
both Hilbert curve and Z-curve.
More precisely, shortest distances are given in:
\begin{itemize}
    \item $13.42\%$ for Hilbert curve,
    \item $12.47\%$ for Z-curve,
    \item $74.04\%$ for H-curve
\end{itemize}
for $1000$ points and $100$ iterations.
In particular, this shows that Hilbert curve is better than Z-curve in $51.8\%$
cases.
This result agrees with the results of~\cite{vukovic2016hilbert}.

\section{Benchmarks}
\label{sec:bench}

Benchmarks were calculated with built-in \texttt{golang} tools on a randomly generated
sets of points given by $32$-bit floating point numbers.
The benchmarks were done~$4$ times for:
\begin{enumerate}
    \item Z-curve only
    \item H-curve,
    \item Hilbert curve,
    \item H-curve with CPU cached static data.
\end{enumerate}
Benchmark results are listed in Table~\ref{tab:bench}.

\begin{table}[ht]
    \centering
    \begin{tabular}{|l|r|}\hline
        Curves & time, ns \\ \hline
        Z & $309$ \\ \hline
        H & $319$ \\ \hline
        Hilbert & $691$ \\ \hline
        H (cached) & $303$ \\ \hline
    \end{tabular}
    \caption{Time for geohashes computation, nanoseconds per computation.}
    \label{tab:bench}
\end{table}

In~\cite{vukovic2016hilbert} benchmarks for the Z-curve only gives
around $490$\,ns per computation (and~$1563$\,ns per computation for both 
Z-curve and Hilbert curve).
Here all the cases for H and/or Hilbert curve contain internally also Z-curve
computation (we have subtracted it to obtain only H-curve and only Hilbert curve).

Benchmarks are done on~\texttt{Intel(R) Core(TM) i9-9900K CPU @ 3.60GHz}.
The main reason of differences of our benchmarks and benchmarks
of~\cite{vukovic2016hilbert} is, more likely, different hardware and, less likely,
some bit different software in building testing environment.

\section{Conclusion}
\label{sec:concl}

We have constructed and implemented geohashing procedure based on usage H-curve.
It appears that H-curve has following advantages:
\begin{itemize}
    \item 
        performance improvement,
    \item in $74\%$ cases lower edit distance for close geographical points
        than for the both Hilbert and Z-curve.
\end{itemize}

This makes H-curve preferable for geohashing than Hilbert curve.

As it was concluded in~\cite{vukovic2016hilbert},
Hilbert curve has some advantages, but it has higher computational complexity, so
practical applications of it are limited.
H-curve gives lower computational overhead and has significantly superior
clustering properties, so applications of H-curve are not so limited as
applications of Hilbert curve.

\section{Acknowledgments}
\label{sec:ackn}

The author is grateful to his Kryptonite colleagues Vasily Dolmatov,
Dr. Nikita Gabdullin and Dr. Anton Raskovalov,
and CEO of PostgresPro Oleg Bartunov for fruitful discussions of topic and results.

\bibliographystyle{unsrt}
\bibliography{refs}

\end{document}